
\magnification=1200

\hyphenation{Di-rac}
\hyphenation{fer-mi-ons}
\rightline{ UdeM-LPN-TH-93-136}
\vskip.5truecm
\centerline{\bf Instantons and the Ground State of the Massive Schwinger Model}
\vskip1truecm
\centerline{M. B. Paranjape$^*$,}
\centerline{Institut f\"ur Theoretische Physik,}
\centerline{Universit\"at Innsbruck, Technikerstrasse 25, Innsbruck, Austria,
A-6020}
\centerline{ and}
\centerline{ Robin Ross,}
\centerline{ Laboratoire de physique nucl\'eaire, D\'epartement de physique}
\centerline{Universit\'e de Montr\'eal, C. P. 6128 succ. ``A"}
\centerline{Montr\'eal, Qu\'ebec, Canada, H3C 3J7}
\vskip1truecm
\centerline{\bf Abstract}
\par\noindent\vskip.5truecm
We study the massive Schwinger model, quantum electrodynamics of massive, Dirac
fermions, in 1+1 dimensions; with space compactified to a circle.  In the limit
that transitions to fermion--anti-fermion pairs can be neglected, we study the
full ground state.  We focus on the effect of instantons which mediate
tunnelling transitions in the induced potential for the dynamical degree of
freedom in the gauge field.
\vskip1truecm
\centerline{\bf I. Introduction and Summary}
\vskip.5truecm\par\noindent
The massive Schwinger model is the direct analog of the quantum electrodynamics
of electrons and positrons in $1+1$ dimensions.  It was studied$^1$ in a
non-perturbative analysis of the effects of the mass term on the phenomena
which
manifest themselves in its massless relative$^2$, the usual  Schwinger model,
quark trapping (confinement) and spontaneous symmetry breaking without massless
scalars (Higg's phenomena).  The usual Schwinger model, is
exactly solvable and has led to much insight concerning the actual structure of
quantum field theories.  It has afforded the first reconciliation
of gauge invariance and the absence of massless states$^2$.  It also provided a
scenario of how confinement could manifest itself at long distances while
asymptotic freedom was valid at short distances$^3$.  The massive model,
further
modified with a current-current interaction was rigorously proven to exist by
Fr\"olich and Seiler$^4$.  Recently, it was shown that its fermionic
determinant
is directly relevant to the fermionic determinant in four dimensional Q.E.D.,
in
the presence of non-constant, though uni-directional electromagnetic
fields$^5$.  Thus the two dimensional model has direct bearing on a physical
four dimensional theory.
\vskip.5truecm\par\noindent
The massive Schwinger model however, is not exactly solvable.  Bosonization
yields a scalar field with a $m\cos (\phi )$ self-interaction term
and an electromagnetic interaction which is no easier to solve.  The mass term
prohibits the possibility of performing chiral gauge transformations, even at
the
classical level, hence the gauge field cannot be removed from the Lagrangean.
Imposition of Gauss' law however, allows for the elimination of all
non-zero momentum modes of the gauge field at the expense of introducing
the Coulomb interaction into the Hamiltonian.  On the circle, we are left with
one dynamical degree of freedom, the Wilson loop of the gauge field.  For the
infinite line the Wilson loop degree of freedom is still there, however it
becomes infinitesimal$^6$.
\vskip.5truecm\par\noindent
Yang-Mills theories defined on a circle have been studied by Manton$^6$
(Schwinger model), by Rajeev$^7$ (pure Yang-Mills) and more recently by
Langmann
and Semenoff$^8$ including massless fermions.  Some of the results in $^8$
pre-empt some of our results, however there are no dynamical calculations done
there, and the normal ordering ground state energy that has been left out there
is very important in the massive case.   Topologically non-trivial gauge
transformations on the circle, render the Wilson loop variable compact, in
fact,
also a circle.  The resulting Hamiltonian contains a kinetic term for the loop
variable, its interaction with the fermions and the properly normal ordered
fermionic kinetic term.  The normal ordering introduces an induced potential
for
the loop degree of freedom.  This potential tends to localize the loop variable
at its minimum.  Instantons, however, mediate tunnelling transitions around the
circle on which the loop variable is defined, which tend to delocalize the loop
variable.
\vskip.5truecm\par\noindent
We focus on the effect of these instantons.  We calculate the semi-classical
correction to the ground state energy.  When these corrections become
appreciable, the spectrum must change from that of a localized particle to that
of an essentially free particle which is constrained to be on a circle.  This
signals the breakdown of the semi-classical limit, however, the corresponding
energy gives an estimate of the energy of the transition regime.  We therefore
find two regimes in the low energy spectrum of the massive Schwinger model.  At
low temperature, the spectrum is essentially a harmonic oscillator spectrum
with
a frequency $\omega$.  This gives a specific heat which is constant for high
temperature, $\beta\omega\rightarrow 0$, but vanishing exponentially for low
temperature,  $\beta\omega\rightarrow\infty$.  Heating the system will
eventually move it into a new regime, where the loop variable delocalizes.
Here
the energy spectrum behaves like $\omega^\prime n^2$, yielding a specific heat
which behaves as $\sqrt{\omega^\prime\over\beta}$ as
$\beta\omega^\prime\rightarrow 0$.  \vskip.5truecm   \centerline{\bf II.
Hamiltonian}  \vskip.5truecm\par\noindent The massive Schwinger model is
governed by the Lagrangean density $$ {\cal L}=-{1\over
4}F_{\mu\nu}(x,t)F^{\mu\nu}(x,t)+\Psi^\dagger    (x,t)(i\partial_t-h(x,t)) \Psi
(x,t), \eqno(1) $$ where $F_{\mu\nu}(x,t)$ is the electromagnetic field
strength
and $h(x,t)$ is the Hamiltonian of the massive Dirac fermion with a minimal
electromagnetic interaction.  We fix the gauge by taking $A_0(x,t)=0$.  This
leaves one gauge field $A_1(x,t)$, and hence $$ F_{01}=\partial_t A_1(x,t)=\dot
v(x,t)=-F_{10}.\eqno(2) $$ The fermionic Hamiltonian is  $$
h(x,t)=-i\gamma_5(\partial_x+iev(x,t))+m\gamma^0,\eqno(3) $$ where
$\gamma_5=i\gamma^0\gamma^1$, and we take the representation
$\gamma_5=\sigma^3$, $\gamma^0=\sigma^1$ in terms of the Pauli matrices.
\vskip.5truecm\par\noindent The equations of motion resulting from the
Lagrangean are Amp\`ere's Law, $$ \ddot v(x,t)=e\Psi^\dagger  (x,t)\gamma_5\Psi
(x,t),\eqno(4) $$ and the Dirac equation, $$
i\partial_t \Psi (x,t)=h(x,t)\Psi (x,t)=\left(-i\gamma_5(\partial_x
+iev(x,t))+m\gamma^0\right) \Psi (x,t),\eqno(5)
$$
Gauss' Law, however, is absent
$$
\partial_x\dot v(x,t)=e\Psi^\dagger   (x,t)\Psi (x,t).\eqno(6)
$$
Invariance of the Lagrangean under static, local gauge transformations
$$
\eqalign{
v(x,t)&\rightarrow v(x,t)-\partial_x\Lambda (x)\cr
\Psi (x,t)&\rightarrow e^{ie\Lambda (x)}\Psi (x,t)}\eqno(7)
$$
yields a local conserved charge ${\cal G}(x)$,
$$
{\cal G}(x)=\partial_x\dot v(x,t)-e\Psi^\dagger   (x,t)\Psi (x,t)\eqno(8)
$$
that is, the time derivative of Gauss' Law is zero,
$$
\partial_t{\cal G}(x)=0.\eqno(9)
$$
\vskip.5truecm\par\noindent
The canonical formalism is straightforward, yeilding the Hamiltonian density
$$
{\cal H}={1\over 2}(\dot v(x))^2+\Psi^\dagger (x,t)h(x,t)\Psi (x,t),\eqno(10)
$$
with canonically conjugate variable pairs
$$
\eqalign{
v(x,t)&,\quad\Pi_{v(x,t)}(x,t)=\dot v(x,t)\cr
\Psi (x,t)&,\quad\Pi_{\Psi(x,t)}(x,t)=i\Psi^\dagger (x,t)}\eqno(11)
$$
and Poisson brackets
$$
\eqalign{
\{ v(x,t_x),\dot v(y,t_y)\}_{t_x=t_y}&=\delta (x-y)\cr
\{ \Psi (x,t_x),i\Psi^\dagger (y,t_y)\}_{t_x=t_y}&=\delta (x-y).}\eqno(12)
$$
\vskip.5truecm\par\noindent
For the quantum theory we work in the Schr\"odinger picture with time
independent operators but time dependent states.  The classical Poisson
brackets are replaced with commutators or anti-commutators,
$$
\eqalign{
\left[ v(x,t_x),\dot v(y,t_y)\right]_{t_x=t_y}&=i\delta (x-y),\cr
\{ \Psi(x,t_x),\Psi^\dagger   (y,t_y)\}_{t_x=t_y}&=\delta (x-y).}\eqno(13)
$$
The quantization proceeds essentially without subtlety, other than operator
ordering ambiguities.  Once the normal ordering infinities are
subtracted, this yields a finite well defined Hamiltonian and Gauss
operator, requiring no regularization.  The normal ordering constants are fixed
by gauge invariance and locality$^9$.
\vskip.5truecm\par\noindent
The fermions can be quantized in the Hilbert space of free, massive Dirac
fermions, while the gauge
fields can be quantized in their corresponding free Hilbert space.  The free
Hamiltonian is
$$
{\hat H}^0=\int dx\left({1\over 2} (\dot {\hat v}(x))^2 + :\Psi^\dagger
(x)h^0(x)\Psi(x):\right)\eqno(14) $$
where
$$
\eqalign{
\Psi (x) &=\sum_{p\in Z}\psi^0_+(x,p)a_p+\psi^0_-(x,p)b_p^\dagger \cr
\hat v(x) &=\sum_{p\in Z}v(p)e^{i{px\over L}}\cr
\dot {\hat v}(x)&=\sum_{p\in Z}{e^{-i{px\over L}}\over 2\pi L}{-id\over
dv(p)}\cr h^0(x) &=-i\gamma^5\partial_x+m\gamma^0 \cr
\psi^0_\pm(x,p)&={e^{i{px\over L}}\over \sqrt{2\pi L}}\psi^0_\pm (p)\cr
\psi^0_\pm (p)&=
{1\over\sqrt{\left(
2(\pm\sqrt{p^2+(mL)^2})(\pm\sqrt{(p)^2+(mL)^2}-p)\right)}}
\pmatrix{(mL)\cr\cr\pm\sqrt{p^2+(mL)^2}-p}}
\eqno(15-20)  $$
and
$$
\{a_p,a^\dagger _q\}=\{b_p,b^\dagger _q\}=\delta_{p,q},\quad{\rm all\quad
others\quad zero.}
\eqno(21)
$$
The normal ordering in $(14)$ is with respect to these operators, the vacuum
state defined by
$$
a_p|0> =b_p|0> =0,\eqno(22)
$$
for all $p$.  Explicitly the free fermionic Hamiltonian is
$$
{\hat H}_F^0={1\over L}\sum_{p\in Z}\sqrt{p^2+(mL)^2}(a_p^\dagger a_p+
b_p^\dagger
b_p).\eqno(23)
$$
The gauge field Hilbert space is presumably a ``wave functional" of the
variables $v(p)$.  This is rather formal here, since the ground state wave
functional would be the infinite fold product of normalized ground state wave
functions for each independent variable, which does not exist.  The
problem arises simply because the infinite fold tensor product is not
isomorphic to the space of normalizable complex valued wave functionals of an
infinite number of variables, with the momentum operator represented by the
functional derivative. This is in contra-distinction to the analogous case for
a
finite number of variables.  Since we will be able to eliminate all but a
single gauge degree of freedom on imposing Gauss' law, we will not overly
concern ourselves with the precise definition of the gauge field Hamiltonian
and Hilbert space.   \vskip.5truecm\par\noindent Introducing the interaction at
the first quantized level yields the Hamiltonian
$$
h(x)=-i\gamma^5(\partial_x+iev(x))+m\gamma^0\eqno(24)
$$
with putative second quantized version
$$
{\hat H}_F={\hat H}_F^0+\int dxv(x)(\Psi^\dagger (x)\gamma^5 \Psi
(x)).\eqno(25) $$
We must define the last term in the R.H.S., the interaction current density
$$
j^1(x)=\Psi^\dagger (x)\gamma^5 \Psi (x)=\bar\Psi (x)\gamma^1 \Psi (x)\eqno(26)
$$
and we will also need the charge density
$$
\rho(x)=\Psi^\dagger (x)\Psi (x)=\bar\Psi (x)\gamma^0 \Psi (x).\eqno(27)
$$
Precisely, we will define,  their corresponding momentum components, for $p\ne
0$ $$
\eqalign{
j^\mu (p) &=\sum_{q\in Z}\left(\langle\psi^0_+(q)|\gamma^0\gamma^\mu
|\psi^0_+(p+q)\rangle a_q^\dagger a_{p+q}+
\langle\psi^0_+(q)|\gamma^0\gamma^\mu
|\psi^0_-(p+q)\rangle a_q^\dagger b_{p+q}^\dagger\right. \cr
&+\left.\langle\psi^0_-(q)|\gamma^0\gamma^\mu |\psi^0_+(p+q)\rangle b_qa_{p+q}-
\langle\psi^0_-(q)|\gamma^0\gamma^\mu |\psi^0_-(p+q)\rangle b_{p+q}^\dagger
b_q\right),}\eqno(28)
$$
where the bracket $\langle\cdot |\cdot\rangle$ is between the spinors in
equation (20).  These are well defined operators with their domain consisting
of states corresponding to finitely many excitations above the free vacuum
state.  For $p=0$ we must actually normal order by subtracting infinite
constants, we take $$
\eqalign{
Q=\int dx j^0(x)&=\sum_{q\in Z}\left(a_q^\dagger a_q-b_q^\dagger b_q\right)\cr
Q_5=\int dx j^1(x)&=\sum_{q\in Z}\left({q\over\sqrt {q^2+(mL)^2}}a_q^\dagger
a_q+
{(mL)\over\sqrt {q^2+(mL)^2}}a_qb_q\right.\cr
&\left.+{(mL)\over\sqrt {q^2+(mL)^2)}}b_q^\dagger
a_q^\dagger -{q\over\sqrt {q^2+(mL)^2)}}b_q^\dagger b_q\right).}\eqno(29)
$$
These charges clearly have arbitrary definitions up to c-numbers, as far as
the fermionic
variables are concerned.  We can fix these c-numbers by imposing gauge
invariance and locality$^9$.
These c-numbers will only affect the fermionic Hamiltonian by c-numbers if we
take equations $(28)$ as the basic building blocks for constructing the
Hamiltonian $(25)$.
\vskip.5truecm\par\noindent
Gauge invariance manifests itself with the condition that
the spectrum of $(25)$ is independent of $v(x)$, we can gauge away  $v(x)$ with
the gauge transformation
$$
e^{ie\int_0^x dyv(y)}.\eqno(30)
$$
Actually we may not remove all of the constant part of $v(x)$, the gauge
transformation $(27)$ must
be single valued on the circle.  Thus we must modify $(30)$ to
$$
e^{ie\int_0^x dy(v(y)-v)}\eqno(31)
$$
with
$$
v={1\over 2\pi L}\int_0^{2\pi L}dxv(x).\eqno(32)
$$
As explained in reference 5, this gives the gauge covariant Hamiltonian
$$
{\hat H}_F={\hat H}_F^0+e\sum_{p\in Z\atop p\ne 0}v(-p)j^1(p)+Le^2
\sum_{p\in Z}v(-p)v(p)+
evQ_5.\eqno(33)
$$
The gauge covariance is manifest if we remove $v(-p)$ dependence by a unitary
transformation $$
{\hat H}_F={\cal U}^\dagger ({\hat H}_F^0+evQ_5+Le^2v^2){\cal U}=
{\cal U}^\dagger {\bar H}_F{\cal U}\eqno(34)
$$
where
$$
{\cal U}=e^{-e\sum_{p\ne 0}{L\over p}v(-p)j^0(p)}.\eqno(35)
$$
This follows from the commutation relations, which can be rigorously
established,
$$
\eqalign{
&[j^0(p),j^1(q)]=2p\delta{p,-q}\cr
&[{\hat H}_F^0,j^0(p)]=-{p\over L}j^1(p)\cr
&[Q,Q_5]=[Q,j^\mu (p)]=[Q_5,j^\mu (p)]=[j^0(p),j^0(q)]=[j^1(p),j^1(q)]=0.}
\eqno(36)
$$
The full Hamiltonian then is
$$
\eqalign{
\hat H&=\sum_{p\in Z}{-1\over 2\pi L}{1\over 2}{d^2\over dv(p)dv(-p)}+
{\hat H}_F\cr
&={\cal U}^\dagger  \left({-1\over 2\pi L}\sum_{p\in Z\atop p\ne 0}{1\over
2}\left({d\over dv(-p)}+{eLj^0(p)\over p}\right)\left({d\over
dv(p)}-{eLj^0(-p)\over p}\right)+{-1\over 4\pi L}{d^2\over dv^2}+{\bar H}_F
\right)
{\cal U}\cr
&={\cal U}^\dagger  \bar H{\cal U}}\eqno(37)
$$
and we are looking for the ground state of $\hat H$ or equivalently $\bar H$.
\vskip.5truecm\par\noindent
We must not forget Gauss' law, which actually simplifies matters.  At the
quantum level we impose Gauss' law as a constraint on physical states.
Physical states are those which are annihilated by the Gauss operator:
$$
\hat{\cal G}(x)|{\rm physical}>=0.\eqno(38)
$$
We can write $\hat{\cal G}(x)$ by Fourier decomposition as
$$
\hat{\cal G}(x)=\sum_{p\in Z}{e^{i{px\over L}}\over 2\pi L}{\hat G}(p)\eqno(39)
$$
where
$$
\eqalign{
{\hat G}(p)&={p\over L}{d\over dv(-p)} -ej^0(p)\cr
&=\cases {{\cal U}^\dagger  {p\over L}{d\over dv(-p)}{\cal U}  &$p\ne 0$\cr
           {\cal U}^\dagger   Q{\cal U}&$p=0$\cr}}\eqno(40)
$$
with $Q$ the charge operator.
\vskip.5truecm\par\noindent
Thus we look for eigenstates
$$
\bar H |{\cal E}> ={\cal E}|{\cal E}>\eqno(41)
$$
subject to the simpler conditions
$$
\cases {{p\over L}{d\over dv(-p)}|{\cal E}>=0 &$p\ne 0$\cr
         Q|{\cal E}>=0 &$p=0$.\cr}\eqno(42)
$$
Clearly the conditions are trivial to satisfy, $|{\cal E}>$ is a charge zero
state, that is indepedent of $v(p)\forall p$.  Then the eigenvalue problem for
$|{\cal E}>$ reduces to
$$
\left({-1\over 2\pi L}\sum_{p\in Z\atop p\ne 0}{1\over
2}e^2L^2{j^0(p)j^0(-p)\over p^2}-{1\over 4\pi L}{d^2\over dv^2}+\bar H_F\right)
|{\cal
E}> ={\cal E}|{\cal E}>.\eqno(43)
$$
The first term is just the Coulomb energy of the fermions while the $v$ is the
only physical degree of freedom in the gauge field.
\vskip.5truecm\par\noindent
We must still deal  with topologically non-trivial gauge transformations
$$
g_k(x)=e^{-i{kx\over  L}}\quad\quad k\in Z.\eqno(44)
$$
The effect on $v$ is to shift it by a constant,
$$
v\rightarrow v+{k\over  eL}.\eqno(45)
$$
Hence $v$ is really a circular variable with circumference ${1\over  eL}$.  The
fermionic part of the Hamiltonian transforms covariantly under these gauge
transformations, the unitary operator $V$ which implements these
transformations
for $k=1$, effects the mapping on the annihilation and creation operators which
diagonalize $\bar H_F$,
$$
\eqalign{
\bar a_p(v)&=V^\dagger \bar a_{p+1}(v-{1\over eL})V\cr
\bar b_p(v)&=V^\dagger \bar b_{p+1}(v-{1\over eL})V.}\eqno(46)
$$
The expression for $V$ is simple,
$$ V=\prod_{p\in Z}e^{-{\pi\over 2}\left(\bar a_p(v)\bar
a_{p+1}^\dagger (v-{1\over eL})-\bar a_{p+1}(v-{1\over eL})\bar a_p^\dagger
(v)\right)}e^{-{\pi\over
2}\left(\bar b_p(v)\bar b_{p+1}^\dagger (v-{1\over eL})-\bar b_{p+1}(v-
{1\over eL})\bar b_p^\dagger
(v)\right)}.\eqno(47)
$$
$V$  is in fact independent of $v$, as can be seen by reexpressing $\bar
a_p(v)$ in
terms of the free annihilation and creation operators,
$$
\eqalign{
&\langle\psi_+(p,v)|\left(|\psi^0_+(p)\rangle a_p+|\psi^0_-(p)\rangle
b^\dagger_p\right)\cr
&= V^\dagger
\langle\psi_+(p+1,v-{1\over
eL})|\left(|\psi^0_+(p+1)\rangle a_{p+1}+|\psi^0_-(p+1)\rangle
b^\dagger_{p+1}\right)V.}\eqno(48)   $$
Therefore, if
$$
\eqalign{ a_p&=V^\dagger
\langle\psi^0_+(p)|\left(|\psi^0_+(p+1)\rangle a_{p+1}+|\psi^0_-(p+1)\rangle
b^\dagger_{p+1}\right)V\cr b^\dagger_p&=V^\dagger
\langle\psi^0_-(p)|\left(|\psi^0_+(p+1)\rangle a_{p+1}+|\psi^0_-(p+1)\rangle
b^\dagger_{p+1}\right)V}\eqno(49)  $$
equation (46) will be satisfied.  This is
a (unitarily implementable) Bogoliubov transformation that is completely
independent of $v$. The Hamiltonian satisfies
$$ V^\dagger \bar H_F(v) V={\hat
H}_F^0+e(v+{1\over  eL})Q_5+e^2L(v+{1\over  eL})^2=\bar  H_F(v+{1\over
eL}).\eqno(50) $$
Thus the spectrum of the fermionic Hamiltonian is invariant after going around
the circle in $v$, however the eigenstates form sections of a bundle over $v$,
the fermionic states satisfy
$$
V|v+{1\over  eL}>=|v>.\eqno(51)
$$
The physical state condition, equation (38), which corresponds to implementing
Gauss' law, makes the states invariant under infinitesimal gauge
transformations.  The integrated version of equation (38), corresponding to
finite gauge transformations that are still continuously connected to the
identity (small), simply implies that the states are invariant under these
small
gauge transformations
$$
e^{i\int dx\Lambda (x){\hat G}(x)}|{\rm physical}>= |{\rm
physical}>.\eqno(52)
$$
In ${\hat G}(x)$, the  part corresponding to the divergence of the electric
field, effects the
transformation on the gauge field, while the fermionic charge density effects
the appropriate
transformation on the fermions.   For the topologically non-trivial gauge
transformations (large), we
can only work with the finite form of these transformations, evidently an
infinitesimal generator
does not exist. We should impose invariance of the states under these
transformations also.  The
transformation on the gauge field is a translation operator, ${\cal K}$,
$$
{\cal K} (v)=v+{1\over eL}\eqno(53)
$$
while the fermions are transformed by $V$, equation (46).  Thus invariance of
the states under large gauge transformations implies
$$
{\cal K} V|{\rm physical}>=|{\rm physical}>.\eqno(54)
$$
The eigenstates of the total Hamiltonian
then must have the form
$$ |{\cal E}>=\sum_n\psi_n(v)|n,v>.\eqno(55) $$
Here $|n,v>$ correspond to a complete set
of fermionic states satisfying equation (51) and $\psi_n(v)$ is the bosonic
wave
function which is periodic under translation of $v$ by ${\cal K}$,
$$
{\cal K} (\psi_n(v))=\psi_n(v+{1\over eL})=\psi_n(v).\eqno(56)
$$
\vskip.5truecm\par\noindent
If we express $\bar H$ in terms of fermionic annihilation and creation
operators which diagonalize $\bar H_F$ we get
$$
\eqalign{
\bar H={-1\over  4\pi L}{d^2\over  dv^2}&+\sum_{p\in Z}\sqrt{({p\over
L}+ev)^2+m^2}\left(\bar a_p^\dagger  (v)\bar a_p(v)+\bar b_p^\dagger
(v)\bar b_p(v)\right)-g(v)\cr
&+{-1\over  2\pi L}\sum_{p\in Z\atop p\ne 0}{1\over
2}e^2L^2{j^0(p)j^0(-p)\over  p^2}}\eqno(57)
$$
and $g(v)$ is the induced potential from the fermions for the gauge degree of
freedom$^9$,
$$
g(v)={-1\over  L}{2mL\over \pi}\sum_{n=1}^\infty{K_1(\pi nmL)\over n}(\cos
(2\pi
neLv)-1).\eqno(58)
$$
\vskip.5truecm
\centerline{\bf III. Approximations}
\vskip.5truecm\par\noindent
The basic approximation that we make, which is not exact$^{10}$, is that
excitations to fermion--anti-fermion pairs (corresponding to $\bar H_F(v)$) are
suppressed.  Standard perturbation theory shows that corrections to the wave
function and energy levels coming from intermediate states are not only
suppressed by explicit factors of some coupling constant, but also due to
powers of the ratio of the characteristic energy scale of the interaction
Hamiltonian with the energy difference between the unperturbed initial state
and the intermediate state$^{11}$.  If the Hamiltonian is a free part plus a
perturbation, $$
H=H^0+\lambda H^\prime\eqno(59)
$$
and the full wave function admits an expansion of the form
$$
\psi=\psi_0+\lambda\sum_{n=1}^\infty\alpha_n\psi_n\eqno(60)
$$
with
$$
H^0\psi_n= E_n\psi_n\quad n=0,1,2,\cdots ,\eqno(61)
$$
we get
$$
\eqalign{
H\psi&=(H^0+\lambda H^\prime)(\psi_0+\lambda\sum_{n=1}^\infty\alpha_n\psi_n)\cr
&=E_0\psi_0+\lambda\sum_{n=1}^\infty\alpha_nE_n\psi_n+
\lambda H^\prime\psi_0+\lambda^2 \sum_{n=1}^\infty\alpha_nH^\prime\psi_n\cr
&=E\psi =E(\psi_0+\lambda\sum_{n=1}^\infty\alpha_n\psi_n).}\eqno(62)
$$
This implies
$$
E=E_0+\lambda <\psi_0|H^\prime |\psi_0>+o(\lambda^2)\eqno(63)
$$
and
$$
\lambda\alpha_n=\lambda{<\psi_n|H^\prime |\psi_0>\over E_0-E_n}+
o(\lambda^2).\eqno(64)
$$
In our case we consider excitations from the fermionic ground state $|0>>$
which is annihilated by $\bar a_p(v)$ and $\bar b_p(v)$.  $j^0(p)$ can be
expressed as a bilinear in these operators, hence the Coulomb energy term
mediates transitions to intermediate states with zero, one or two
fermion--anti-fermion pairs,  at zero total momentum.  Then
$$
\lambda H^\prime ={(eL)^2\over 4\pi}\left({1\over L}\sum_{p\in Z\atop p\ne 0}
{j^0(p)j^0(-p)\over p^2}\right),\eqno(65)
$$
and the coefficients of states involving fermion--anti-fermion pairs, first
order in the perturbative expansion of the full ground state are
$$
\eqalign{
&{(eL)^2\over 4\pi}{\left({1\over L}\sum_{p\in Z\atop p\ne 0}
{<{\rm pairs}|j^0(p)j^0(-p)|0>>\over p^2}\right)\over (E_0-E_{\rm pairs})}\cr
&\le {(eL)^2\over 4\pi 2mL}\left({1\over L}\sum_{p\in Z\atop p\ne 0}
{<{\rm pairs}|j^0(p)j^0(-p)|0>>\over p^2}\right).}\eqno(66)
$$
This behaviour continues in each order.
Thus the coupling constant emerges as $(eL)({e\over m})$, which we take to be
arbitrarily small.  Therefore we can neglect the contribution of pair
states, arising because of the Coulomb term, to the full ground state.  This
makes intuitive sense in the following way: on a circle, we cannot separate
charges to infinity, there is a maximum separation that we can separate
charges.  The Coulomb energy, which is linear in the separation, is bounded.
It
is a little more work to see that the corresponding operator is relatively
bounded in comparison to the fermionic Hamiltonian about which we perturb.
Hence in the limit that its coefficient goes to zero, we can rigorously neglect
it.  \vskip.5truecm\par\noindent The shift in the ground state energy has the
leading contribution
$$
\eqalign{
&<<0|\lambda H^\prime |0>>\cr
&={(eL)^2\over 4\pi}\left({1\over L}\sum_{p\in Z\atop
p\ne 0} {<<0|j^0(p)j^0(-p)|0>>\over p^2}\right)\cr
&={(eL)^2\over 4\pi}\left({1\over L}\sum_{p\in Z\atop
p\ne 0} {\sum_{q\in Z}\big|\langle\psi_-(q,v)\big|\psi_+(p+q,v)\rangle\big|^2
\over
p^2}\right)\cr
&={(eL)^2\over 4\pi}{1\over L}\sum_{p,q\in Z\atop
p\ne 0}
{1\over p^2}\left({1\over
2\sqrt{(q+eLv)^2+(mL)^2}(\sqrt{(q+eLv)^2+(mL)^2}+q+eLv)
}\right)\times\cr
&\left((mL)^2\right.\cr
&\left. -(\sqrt{(q+eLv)^2+(mL)^2}+q+eLv)(\sqrt{(p+q+eLv)^2+(mL)^2}-(p+q+eLv))
\right)^2\times\cr
&\left({1\over 2\sqrt{(p+q+eLv)^2+(mL)^2}(\sqrt{(p+q+eLv)^2+(mL)^2}-(p+q+eLv))}
\right).}\eqno(67)
$$
This has as coefficient $(eL)^2$ multiplying a function of $eLv$, which is of
order $1$.  This shift in the ground state energy is suppressed relative
to $-g(v)$, the normal ordering contribution to the ground state energy, by
again a factor of  $(eL)({e\over m})$.
\vskip.5truecm\par\noindent
This yields the truncated Hamiltonian
$$
H={-1\over  4\pi L}{d^2\over  dv^2}+{1\over L}\sum_{p\in
Z}\sqrt{(p+ev)^2+(mL)^2}\left(\bar a_p^\dagger  (v)\bar a_p(v)+\bar b_p^\dagger
(v)\bar b_p(v)\right)-g(v)\eqno(68)
$$
and the ground state will be of the form
$$
|{\cal E}_0>=\psi (v)|0>>.\eqno(69)
$$
The state $|0>>$ depends on $v$, hence the derivative ${d\over  dv}$ will give
transitions to other fermionic states.  We have
$$
|0>>={\cal W}^\dagger |0>\eqno(70)
$$
with $|0>$ the free fermion vacuum, and
$$
{\cal W}=\prod_{p\in Z}e^{\theta
(p,v)(a^\dagger_pb^\dagger_p-b_pa_p).}\eqno(71)
$$
${\cal W}$ is the unitary operator implementing the Bogoliubov transformation
$$
\eqalign{
\bar a_p(v)&=\cos(\theta (p,v))a_p+\sin(\theta (p,v))b^\dagger_p={\cal
W}^\dagger a_p{\cal W}\cr
\bar b^\dagger_p(v)&=-\sin(\theta (p,v))a_p+\cos(\theta
(p,v))b^\dagger_p={\cal W}^\dagger b^\dagger_p{\cal W},} \eqno(72)
$$
with
$$
\eqalign{
\cos(\theta (p,v))&=<\psi_+(p,v)|\psi^0_+(p)>\cr
\sin(\theta (p,v))&=<\psi_+(p,v)|\psi^0_-(p)>.}\eqno(73)
$$
Then,
$$
{d\over  dv}|0>>=-\sum_{p\in Z}{d(\theta (p,v))\over  dv}\bar
a^\dagger_p(v)\bar
b^\dagger_p(v)|0>>.\eqno(74)
$$
The derivative can be expressed as
$$
{d(\theta (p,v))\over  dv}={1\over \cos(\theta (p,v))}{d\sin(\theta (p,v))\over
dv}= {1\over <\psi_+(p,v)|\psi^0_+(p)>}{d<\psi_+(p,v)|\psi^0_-(p)>\over
dv}.\eqno(75)
$$
The denominator is a smooth function of order $1$ in $eLv$, and also in $mL$.
Note that $eLv$ is always in $[0,1]$.  The numerator is
$$
\eqalign{
&\langle\psi_+(p,v)|\psi^0_-(p)\rangle =\cr
&{(mL)^2-(\sqrt{(p+eLv)^2+(mL)^2}-(p+eLv))(\sqrt{p^2+(mL)^2}+p)\over
2\sqrt{\sqrt{(p+eLv)^2+(mL)^2}(\sqrt{(p+eLv)^2+(mL)^2}-(p+eLv))
}}\times\cr
&\times{1\over\sqrt{\sqrt{p^2+(mL)^2}(\sqrt{p^2+(mL)^2}+p)}}}\eqno(76)
$$
It is easy to see that $<\psi_+(p,v)|\psi^0_-(p)>$ behaves like ${eLv\over mL}$
for $mL>>p$.  For $mL<<p$ it behaves like
${mLeLv\over p^2}$, which is then much smaller than ${eLv\over mL}$ and for
$mL=\alpha p$, with  $\alpha\approx 1$, we can directly factor the powers of
$mL$ out of the expression leaving a function of ${eLv\over mL}$ and $\alpha$.
Thus differentiating with respect to $v$ in each case gives a factor of
${eL\over mL}$.  Thus when this
parameter is small we can neglect the action of the derivative ${d\over dv}$ on
the fermionic state, yielding the equation for $\psi (v)$
$$
({-1\over  4\pi L}{d^2\over  dv^2}-g(v))\psi (v)={\cal E}^0\psi (v),\eqno(77)
$$
a simple, one-dimensional quantum mechanics problem on a circle with a periodic
potential.  For $m\ne 0$, $g(v)$ is a smooth periodic potential with a single,
symmetric well at $v={1\over  2eL}$.  It is clear what the excitation
spectrum will
be.  For low energies, the variable will be localized in the bottom of the
approximately harmonic well.  The energy will be
$$
{\cal E}_n = \hbar \omega (n+{1\over  2})\quad n=0,1,2,\cdots ,\eqno(78)
$$
where $\omega$ is the curvature at the bottom of the well.  Then for high
energies,
the variable $v$ will hardly notice the small potential $-g(v)$, but will be
essentially constrained by the size of the circle upon which it must sit.  The
circumference is ${1\over  eL}$, giving rise to an energy spectrum
$$
{\cal E}_n =\hbar \omega^\prime (n^\prime )^2\quad n^\prime = 1,2,\cdots .
\eqno(79)
$$
$\omega $ is given by
$$
\omega = \sqrt {{1\over  2\pi L}\left({d^2\over  dv^2}g(v)\Big|_{v={1\over
2eL}}\right)},\eqno(80)
$$
while
$$
\omega^\prime = \pi e^2 L.\eqno(81)
$$
\vskip.5truecm
\centerline{\bf IV.  Instantons}
\vskip.5truecm\noindent
We proceed along the lines pioneered by Langer$^{12}$, and popularized by
Coleman$^{13}$, for using the Euclidean path integral to calculate the effects
of
tunnelling.  The idea is simple, the matrix element of $e^{-TH}$ in the
``position" eigenstate $|v={1\over 2eL}>$ has a representation in terms
of a Euclidean path integral
$$
< v={1\over 2eL}|e^{-TH}|v={1\over 2eL}> =\int_{v({-T\over 2})=
v({T\over 2})={1\over 2eL}}{\cal D}v(\tau )e^{-\int_{-T\over 4\pi L}^{T\over
4\pi L}d\tau L^E(v(\tau ))},\eqno(82)
$$
where $L^E(v(\tau ))$ is the continuation to (dimensionless) Euclidean time
$\tau$ of the usual Lagrangean.  In this simple case it corresponds to
$$
L^E(v(\tau ))={1\over 2}(\dot v(\tau))^2 -2\pi Lg(v(\tau))\eqno(83)
$$
which can be thought of simply as the Lagrangean describing motion of a
particle in minus the original potential, $-(-g(v))$.  The matrix element
equation $(82)$ has the expansion
$$
< v={1\over 2eL}|e^{-TH}|v={1\over 2eL}> =e^{-T{\cal E}_0}
< v={1\over 2eL}|{\cal E}_0>< {\cal E}_0|v={1\over 2eL}> + \cdots\eqno(84)
$$
thus in the limit that $T\rightarrow\infty$ we can extract ${\cal E}_0$, and
the
amplitudes $|< {\cal E}_0|v={1\over 2eL}> |^2$; contributions from higher
states will be exponentially suppressed.
\vskip.5truecm\par\noindent
The Euclidean functional integral
can be evaluated in a saddle point approximation.  The first step is to
identify the saddle point, called an instanton here, and then perform the
functional integral in a Gaussian approximation about the saddle point.
The Gaussian functional integral simply gives rise to $e^{-S_0^E}$ where
$S_0^E$
is the Euclidean action for an instanton, multiplied by the inverse square root
of a functional determinant.  The determinant is the product of all
the eigenvalues of the functional operator corresponding to the second
variation of the Euclidean action about the instanton.  This product has
two main problems, it is of course infinite, but it is also zero!
\vskip.5truecm\par\noindent
The infinite product of a continuum of eigenvalues that become arbitrarily
large is formally infinite, but in fact completely ill-defined.  It is,
however, only the ratio of this product relative to the correponding
(infinite) product for the free case that is important.  This ratio is
finite.  The other problem comes from vanishing eigenvalues.  These render the
determinant zero.  Such zero modes correspond to degeneracies of the original
instanton.  There usually exist a whole set of instantons with the same
action.  We should sum over the contribution from all saddle points
(instantons) with the same, minimal action.  When we perform this sum, we have
already taken into account the direction in function space corresponding to
the zero modes.  Thus in the Gaussian integral, we should exclude the
integration along the zero modes, the result being the determinant with the
zero eigenvalues removed. There is a Jacobian factor which must be taken into
account since the measure corresponding to summing over the contribution from
degenerate instantons is different from that corresponding to integrating over
the zero modes directions in the Gaussian functional integral.
\vskip.5truecm\par\noindent
Actually for large but finite $T$, there are no exact zero modes,
corresponding to invariance under translation of the instanton in Euclidean
time.  The corresponding eigenvalue however, is exponentially small in $T$,
thus the infinite $T$ calculations will be exponentially close to those for
finite, but large $T$.  Furthermore, we must recognize that in this case,
there are other approximate critical points, corresponding to $N$ widely
separated instantons which must also be considered. The
corresponding action is $N$ times the action of one instanton, implying
naively that their contribution is suppressed by $N-1$ powers of $e^{-S_0^E}$
relative to the contribution for one instanton.  The degeneracy factor of these
approximate critical points is, however, ${T^N\over N!}$, corresponding to
independent translation in Euclidean time of each instanton.  This factor can
be arbitrarily large compensating the suppression from the exponential factor,
until $N$ surpasses $T$.  $T$ must always of course be sufficiently large so
that the space per instanton, $T\over N$, is still much larger that the size
of the instanton.  The size of the instanton is determined by the parameters
that appear in the Lagrangean, hence has nothing to do with $T$ and $N$.  Thus
it is always possible to satisfy this constraint.  We should sum over $N$ until
it is of the same order as $T$.  However, once $N$ is of this size, the
contribution of further terms in the expansion is exponentially small, due to
the $1\over N!$, thus we make only a negligible error to continue the sum up
to $\infty$.
\vskip.5truecm\par\noindent                                        For $mL$
sufficiently large we can keep only the first term in the series for $g(v)$,
$$
-g(v)\rightarrow {-4mL\over  \pi L}K_1(2\pi m L) \sin^2(\pi eLv)\eqno(85)
$$
and
$$
\omega = 4\pi eL\sqrt{mLK_1(2\pi m
L)},\eqno(86)
$$
is the relevant frequency for a (dimensionless) Euclidean time, $\tau$.
The corresponding instanton equation, obtained by varying the
Euclidean Lagrangean is
$$ {d^2\over  d\tau^2}\bar
v(\tau)=-8\pi eLmLK_1(2\pi m L)\sin (2\pi eL\bar v(\tau ) ).\eqno(87)
$$
This equation is easily integrated to give
$$
\bar v(\tau )={2\over  \pi e L}\tan^{-1}\left(e^{\pm \omega (\tau -\tau_0
)}\right)-{1\over  2eL},\eqno(88)
$$
the $\pm$ choosing an instanton or an anti-instanton.  The action for either
is given by
$$
S_0= {8\sqrt{mLK_1(2\pi m L)}\over  \pi e L}.\eqno(89)
$$
\vskip.5truecm\par\noindent
A method which we follow here, for calculating the ratio ${{\rm
det}^\prime\over {\rm det}_0}$, where ${\rm
det}^\prime$ is the determinant with the zero modes excluded and ${\rm
det}_0$ is the free determinant, is given in Coleman$^{13}$.  Here it is
shown that  $$
{{\rm det}^\prime\over {\rm det}_0}=
{\psi_0({T\over 4\pi L})\over\lambda_0\psi^0_0({T\over 4\pi L})}\eqno(90) $$
where $\psi_0({T\over 4\pi L})$ is the eigenfunction (evaluated at
${T\over 4\pi L}$) with the smallest eigenvalue $\lambda_0$,  for the
differential equation corresponding to the Schr\"odinger operator on the
interval
$[-{T\over 4\pi L},{T\over 4\pi L}]$, with potential $+g({\bar v}(\tau
))^{\prime\prime}$ satisfying the boundary conditions $\psi_0(-{T\over
2})=0$, $\psi_0^\prime (-{T\over 4\pi L})=1$.  $\psi^0_0({T\over 4\pi L})$ is
the
analogous solution for the free problem.  As $T\rightarrow\infty$, these are
easy to find.  We find
$$
{{\rm det}^\prime\over {\rm det}_0}={S_0(\pi eL)^2\over 8\omega^3} ={1\over
4\omega^2}.\eqno(91)
$$
The contribution from the instantons and the
anti-instantons sums separately.  There is no constraint on the order in which
they must appear, either tunnels from the same initial and final state.  This
gives two times the same contribution.  The factor$^{13}$ $K$, which takes into
account the Jacobian factor and the ratio of the determinant in the presence of
one instanton and the free determinant then is
$$
K=\left({S_0\over 2\pi}\right)^{1\over 2}\left({{\rm det}^\prime\over {\rm
det}_0}\right)^{-{1\over 2}}=({\omega\over \pi})^{1\over 2}{2\omega\over
eL\pi}.
\eqno(92) $$
Then we find the path integral, equation
(82), is given by (in dimensionless Euclidean time)
$$
\left({\rm det}_0\right)\sum_{N_1,N_2=0}^\infty
{({T\over 2\pi L}Ke^{S_0})^{N_1}({T\over 2\pi L}Ke^{S_0})^{N_2}\over N_1!N_2!}=
({\omega\over
\pi})^{1\over  2}e^{-{T\over 2\pi L}{1\over  2}\omega}e^{2{T\over 2\pi
L}Ke^{-S_0}}.\eqno(93)
$$
The sum over $N_1$ is for the instantons while that over $N_2$ is for the
anti-instantons, the upshot is the factor of $2$ in the exponent.  The free
determinant is calculated in Coleman$^{13}$.
\vskip.5truecm\par\noindent
Finally we find,
$$
\eqalign{
<v={1\over  2eL}|e^{-TH}|v={1\over  2eL}>&=e^{-T{\cal E}_0}<v={1\over
2eL}|{\cal
E}_0><{\cal E}_0|v={1\over  2eL}>+\cdots\cr
&=({\omega\over \pi})^{1\over  2}e^{-{T\over  2\pi L}({1\over
2}\omega -2({\omega\over \pi})^{1\over 2}{2\omega\over \pi
eL}e^{-S_0})}.}\eqno(94)
$$
Hence
$$
\eqalign{
{\cal E}_0&={1\over  2 \pi L}{1\over  2}\omega
\left(1-{16\over \pi}\left({\sqrt{mLK_1(2\pi mL)}\over  eL}\right)^{1\over
2}e^{-S_0}\right)-{1\over  2 \pi L}8mLK_1(2\pi mL)\cr
&={1\over  2 \pi L}{1\over  2}\omega\left(1-{4\sqrt{2}\over
\sqrt{\pi}}\sqrt{S_0}e^{-S_0}\right)-{1\over  2 \pi L}8mLK_1(2\pi
mL),}\eqno(95)
$$
where we have included the offset due to the value of the minimum of the
potential
$-g(v)$.
\vskip.5truecm\par\noindent
The correction that we have calculated is non-perturbative.  Our
approximations, $eL{e\over  m}\rightarrow 0$, ${e\over  m}\rightarrow 0$ and
$mL\rightarrow\infty$ still leave $S_0$ arbitrary (as can be seen from its
expression).  It can take values from $0$ to $\infty$.  The function
$e^{-x}\sqrt x$ is maximum at $x={1\over  2}$, where it is equal to ${1\over
\sqrt{2e}}$.  Then ${4\sqrt{2}\over  \sqrt\pi}{1\over \sqrt{2e}} \cong 1.3$.
Hence as $S_0$ approaches $1\over  2$ from above or below, the effects of the
instantons become non-negligible.
\vskip.5truecm\par\noindent
This work supported in part by NSERC of Canada and by FCAR du Qu\'ebec and the
Universit\"at Innsbruck, Naturwissenschaftliche Fakultaet.  We thank G. Gr\"ubl
and F. Ehlotzky for making our visit to Innsbruck possible.  We also thank G.
Gr\"ubl, J. LeTourneux, R. MacKenzie and N. Manton for useful discussions.
\vskip1truecm
\centerline{References}
\vskip1truecm
\par\noindent
$^*$ permanent address, Laboratoire de physique nucl\'eaire, Universit\'e de
Montr\'eal, C.P. 6128 succ. ``A", Montr\'eal, Qu\'ebec, Canada, H3c 3J7.
\par\noindent
1.  S. Coleman, R. Jackiw, L. Susskind, Annal of Physics, {\bf 93},267,
(1975);S. Coleman, ibid, {\bf 101}, 239, (1976).
\par\noindent
2.  J. Schwinger, Phys. Rev., {\bf 125}, 397, (1962);ibid, {\bf 128}, 2425,
(1962).
\par\noindent
3.  J. Kogut, L. Susskind, Phys. Rev., {\bf D10}, 732, (1974).
\par\noindent
4.  J. Fr\"ohlich and E. Seiler, Helvetica Physica Acta, {\bf 49}, 889, (1976).
\par\noindent
5.  E. M. Fry, Phys. Rev., {\bf D45}, 682, (1992);Dublin preprint, (1992).
\par\noindent
6.  N. S. Manton, Annals of Physics, {\bf 159}, 220, (1985).
\par\noindent
7.  S. Rajeev, Phys. Lett., {\bf B212}, 203, (1988).
\par\noindent
8.  E. Langmann and G. Semenoff, Physics Lett., {\bf B296},117, (1992).
\par\noindent
9.  M. B. Paranjape, Phys. Rev., {\bf D40}, 540, (1989).
\par\noindent
10.  A. Bohm, Proceedings of the 23 GIFT/Nato ASI,Salamanca,Spain,June 1992.
\par\noindent
11. L. Schiff, Quantum Mechanics, McGraw-Hill, New York, (1949).
\par\noindent
12.  J. S. Langer, Annals of Physics, {\bf 41}, 108, (1967).
\par\noindent
13.  S. Coleman, ``Uses of Instantons", Erice lecture notes, in ``The Whys of
Sub-nuclear Physics", Plenum, (1979).
\end